\documentclass[sigconf,nonacm]{acmart}
\AtBeginDocument{%
  }

\setcopyright{none}

\usepackage{xcolor}
\usepackage{enumitem}
\usepackage{tabularx}
\usepackage{booktabs}
\usepackage{multirow}
\newcommand{\jiajun}[1]{\textcolor{black}{#1}}

\begin{document}

\title{\jiajun{LogicTree: Object-Oriented Writing for Structured Human-AI Academic Authoring}}

\author{Jiajun Zhu}
\affiliation{%
  \institution{The Hong Kong University of Science and Technology}
  \city{Hong Kong}
  \country{China}
}
\email{jzhucq@connect.ust.hk}

\author{Xinyu Cheng}
\affiliation{%
  \institution{Peking University}
  \city{Beijing}
  \country{China}
}
\email{chengxinyu26@stu.pku.edu.cn}

\author{Wenshuo Zhang}
\affiliation{%
  \institution{The Hong Kong University of Science and Technology}
  \city{Hong Kong}
  \country{China}
}
\email{wzhangeb@connect.ust.hk}

\author{Huamin Qu}
\affiliation{%
  \institution{The Hong Kong University of Science and Technology}
  \city{Hong Kong}
  \country{China}
}
\email{huamin@cse.ust.hk}

\author{Yanna Lin}
\authornote{Corresponding author.}
\affiliation{%
  \institution{University of Waterloo}
  \city{Waterloo}
  \state{Ontario}
  \country{Canada}
}
\email{yanna.lin@uwaterloo.ca}

\renewcommand{\shortauthors}{Zhu et al.}
\hypersetup{pdfauthor={Jiajun Zhu, Xinyu Cheng, Wenshuo Zhang, Huamin Qu, Yanna Lin}}
\newcommand{\yanna}[1]{\textcolor{black}{#1}}
\renewcommand{\figureautorefname}{Fig.}
\renewcommand{\tableautorefname}{Tab.}
\renewcommand{\sectionautorefname}{Sec.}
\renewcommand{\subsectionautorefname}{Sec.}
\renewcommand{\subsubsectionautorefname}{Sec.}

\begin{abstract}
Academic writing is non-linear, iterative, and structure-driven, requiring authors to develop fragmented ideas into arguments with local coherence and global consistency.
\jiajun{However, current AI-assisted tools primarily exchange linear prompts and free-form text, forcing authors to reconstruct generated logic and repeatedly verify its alignment with their structural intentions.}
\jiajun{We propose \textit{object-oriented writing} (OOW), an OOD-inspired representation and interaction paradigm in which persistent writing objects serve as shared, manipulable referents for authors and AI.}
\jiajun{We instantiate OOW in LogicTree, an Overleaf-integrated prototype that combines a hierarchical object representation with a structure-grounded writing assistance module for author-initiated generation and issue-grounded revision and a structural logical review module for inspecting local coherence and global consistency.}
\jiajun{A within-subjects study (N=12) provides initial evidence that, in controlled academic writing tasks, LogicTree helps users structure ideas, reduces workload, and improves measured logical quality compared with a conversational baseline.}
\end{abstract}



\begin{teaserfigure}
    \centering
    \includegraphics[width=\textwidth]{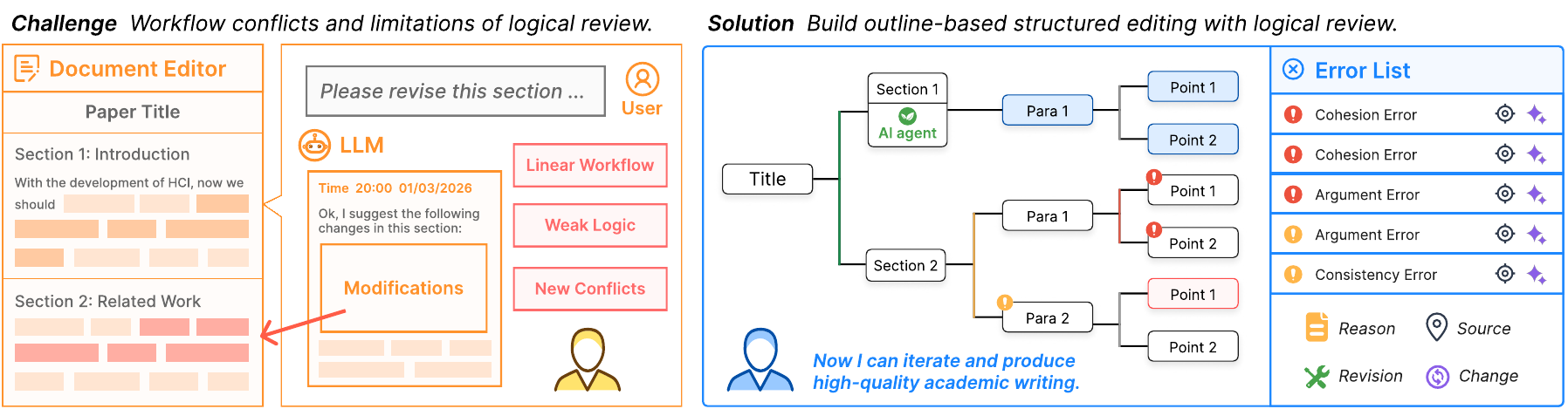}
    \caption{\jiajun{LogicTree instantiates \textit{object-oriented writing} as a persistent hierarchical object representation shared by authors and two computational modules for writing assistance and structural logical review.}}
    \Description{A side-by-side illustration contrasts a linear document-editor and chat-based LLM workflow with LogicTree. The linear workflow can produce weak logic and new conflicts. LogicTree represents the document as a hierarchy of title, sections, paragraphs, and points, supports AI-assisted expansion from selected objects, marks potential cohesion and argument issues in the tree, and presents corresponding reasons, sources, revisions, and changes in an error list.}
    \label{fig:teaser}
\end{teaserfigure}


\maketitle

\raggedbottom
\setlength{\abovecaptionskip}{8pt}   
\setlength{\textfloatsep}{8pt}   

\section{Introduction}

Academic writing is a fundamental yet cognitively intensive activity for researchers across disciplines \cite{ncte:/content/journals/10.58680/ccc198115885,kellogg2008training}.
Authors do not simply compose text from start to finish. 
They progressively develop fragmented ideas, refine outlines, reorganize arguments, and maintain logical coherence across multiple revisions. 
This process demands rigorous reasoning, clear argument, and consistent evidential support. 
It remains one of the most time-consuming and error-prone stages of academic work \cite{daiute1986physical, kellogg2013model}.

Large language models (LLMs) have been widely adopted to assist academic writing. 
They excel at text generation, grammar correction, and stylistic polishing, lowering the barrier to producing fluent text \cite{zhao2026surveylargelanguagemodels, Lin_2024}. 
However, when applied to long-form and logic-intensive academic writing, existing AI-assisted tools exhibit fundamental limitations. 
They typically treat writing as a linear process of generating or revising text in a conversational manner \cite{10.1145/3491102.3502030, 10.1145/3490099.3511105}. 
As a result, users receive dense blocks of generated text and must work backward to infer, reconstruct, and verify the underlying logic. 
In contrast, academic writing is inherently non-linear and structure-centered. 
Authors start from fragmented ideas, iteratively build hierarchical arguments, and continuously adjust structure while filling in content. 
This natural workflow clashes with the linear output of current AI-assisted tools, causing a gradual misalignment between what authors intend to argue and what the generated text actually conveys \cite{mysore-etal-2025-prototypical, betley2026training}. 
This mismatch creates two core difficulties. 
On one hand, users struggle to understand and verify AI outputs because argument structure and logical dependencies are embedded within linear text, requiring manual extraction and reconstruction. 
On the other hand, conveying writing intentions is equally challenging, as authors think in terms of structural elements such as section purposes and argument hierarchies, yet have to translate these into linear prompts or local edits, constantly checking whether the generated text aligns with their goals.

Prior work has pursued two main directions to address these difficulties. 
One direction uses structured representations such as outlines \cite{engelbart1962augmenting, notion} or node-graph interfaces \cite{10.1145/3586183.3606737, 10.1145/3613904.3642400} to organize content in a non-linear format, aiming to reduce the burden of understanding linear text. 
However, these systems treat structure primarily as a navigation tool, not as a manipulable network of logical relationships. 
Another direction focuses on helping users articulate their intentions more clearly through conversational prompts \cite{10.1145/3613904.3642217, 10.1145/3586183.3606756} or goal articulation. 
Nevertheless, these tools still operate within a linear prompt-to-text framework, offering no support for tracking or revising structural intentions across the long writing process. 
More recently, some work has explored introducing a shared intermediate representation between humans and AI to improve collaboration and alignment \cite{10.1145/3491102.3517582, 10.1145/3654777.3676357, 10.1145/3746059.3747668}, suggesting the problem lies in the lack of an operable representation layer. 
However, existing approaches are designed for one-shot or short-cycle tasks, and do not support the long-term evolution, structural reorganization, and cross-section logic maintenance required in academic writing.

To better understand how these limitations manifest in real academic writing workflows, we conducted a formative study with six doctoral students and researchers (\autoref{Section 3}). 
Participants described a tightly coupled, highly iterative workflow of drafting, reviewing, and revising. 
They reported that current AI tools are useful for grammar correction and wording refinement, but struggle to support the progressive organization of fragmented ideas into structured arguments (\textbf{C1}) and the inspection of logical issues, such as local coherence and global consistency (\textbf{C2}). 
These findings confirmed the requirement for a framework that externalizes logical structure, makes it inspectable, and enables controllable AI assistance.

\begin{figure}[t]
    \centering
    \includegraphics[width=\linewidth]{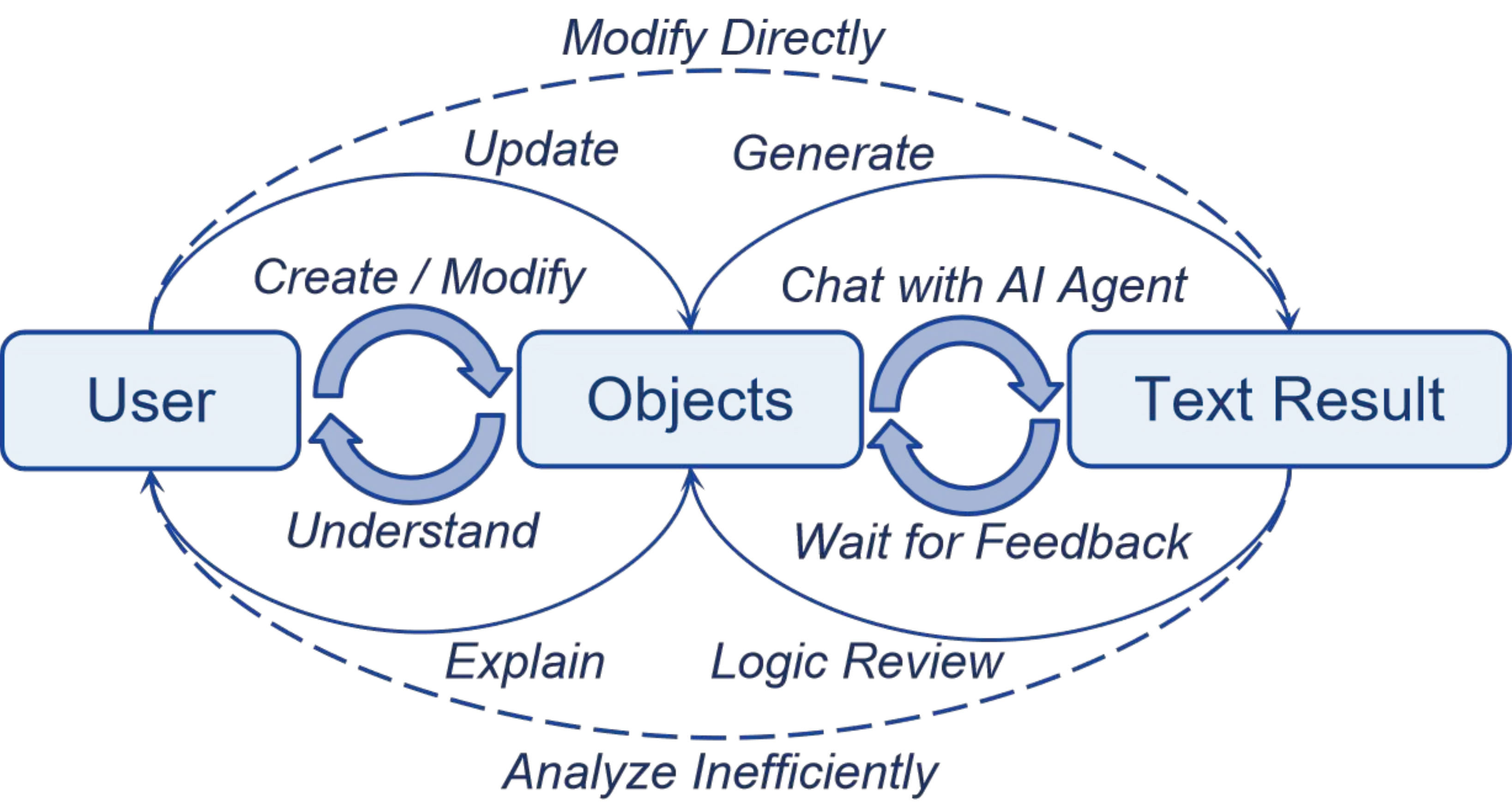}
    \caption{\jiajun{\textit{Object-oriented writing} introduces persistent writing objects as shared, manipulable referents between authors and AI.}
    \jiajun{LogicTree instantiates the paradigm with a hierarchical object representation and computational modules for structure-grounded writing assistance and structural logical review that operate over the same evolving structure.}}
    \Description{\jiajun{A comparison between direct linear text editing and \textit{object-oriented writing}.}
    \jiajun{LogicTree represents writing as container and propositional objects and connects object-grounded writing assistance with logical review in a user-controlled iterative loop.}}
    \label{fig:concept}
\end{figure}

Based on these insights, we propose \textit{object-oriented writing} (\jiajun{OOW;} \autoref{fig:concept}), \jiajun{an OOD-inspired representation and interaction paradigm in which identifiable academic-writing units persist as shared, manipulable objects for authors and AI}.
\jiajun{Each writing object maintains explicit state, supports scoped operations, and participates in structural or rhetorical relations with other objects.}
\jiajun{This intermediate representation externalizes writing intentions and logical structure so that authors and AI can refer to the same units rather than communicating solely through linear prompts and generated prose.}

We \jiajun{instantiate OOW} in \jiajun{LogicTree}, a prototype integrated into Overleaf.
\jiajun{LogicTree represents sections and paragraphs as container objects that organize scope and context, while leaf-level propositional objects additionally carry argument roles and participate in logical relations.}
\jiajun{It couples two complementary computational modules: a structure-grounded writing assistance module that supports author-initiated generation and issue-grounded revision, and a structural logical review module that inspects linear cohesion, argument coherence, and global consistency.}
\jiajun{The review module is guided by linguistic theories \cite{hasan1984coherence, halliday2014cohesion, toulmin2003uses, rst} and presents its results as inspectable, user-contestable feedback rather than binding judgments.}
A within-subjects user study (N=12) compared the \jiajun{integrated LogicTree workflow} against a conversational baseline.
\jiajun{In these controlled academic writing tasks, LogicTree} \jiajun{reduced} workload and \jiajun{received higher ratings and logical-quality scores} across \jiajun{the} measured dimensions.

In summary, our contributions are three-fold:
\begin{itemize}[leftmargin=*, labelindent=0pt, itemindent=0pt]
    \item \jiajun{We introduce \textit{object-oriented writing}, an OOD-inspired representation and interaction paradigm that uses persistent writing objects as shared referents between authors and AI.}
    \item \jiajun{We present LogicTree, an Overleaf-integrated instantiation of OOW that combines a hierarchical object representation with structure-grounded writing assistance and structural logical review modules while preserving user control.}
    \item \jiajun{We provide initial empirical evidence from a within-subjects user study (N=12) that compares the integrated LogicTree workflow with a conversational baseline in controlled academic writing tasks.}
\end{itemize}

\section{Related Work}
We review AI-assisted academic writing and structured content manipulation, which are closely related to our study.

\subsection{AI-Assisted Academic Writing}
Academic writing 
demands rigorous logical reasoning, well-structured argument, and clear evidential support, placing high 
cognitive and professional demands on authors \cite{swales_genre_1990, ncte:/content/journals/10.58680/ccc198115885}. 
With the development of large language models (LLMs), AI-assisted tools are increasingly used for academic writing \cite{10.1016/j.ijinfomgt.2023.102642}. 
They excel at text generation, language polishing, and surface-level grammar checking, lowering the barrier to
fluent and well-formed text \cite{zhao2026surveylargelanguagemodels, Lin_2024}. 
However, in long-form academic writing, these tools are constrained by known LLM limitations such as hallucination \cite{10.1145/3571730} and the ``lost in the middle'' phenomenon \cite{liu-etal-2024-lost}. 
Moreover, HCI researchers have noted that these systems often overlook the recursive nature of scholarly thought \cite{10.1145/3613904.3642400}, hindering user intent alignment \cite{10.1145/3442188.3445922, 10.1145/3571730}. 
These limitations motivate further research into AI-assisted academic writing.

Current AI-assisted writing tools treat writing as a linear process of text generation and local revision. 
Systems such as CoAuthor \cite{10.1145/3491102.3502030} and Wordcraft \cite{10.1145/3490099.3511105} provide sentence or paragraph level continuation, while commercial tools like Grammarly \cite{grammarly} and Writefull \cite{writefull} focus on localized grammar and style correction. 
This sentence-level granularity is insufficient for managing logical dependencies across a manuscript.
More recent research attempts to raise the abstraction level by operating on user intentions. 
For example, Rambler \cite{10.1145/3613904.3642217} helps users articulate writing goals through conversational prompts, and Sensecape \cite{10.1145/3586183.3606756} supports hierarchical organization of information needs. 
However, both still return a single piece of text as output, without modeling cross-section relationships or supporting interactive manipulation of the underlying logic. 
While works like Graphologue \cite{10.1145/3586183.3606737} and Luminate \cite{10.1145/3613904.3642400} introduce structured abstractions such as node-graph representations, they treat the manuscript as a static collection of blocks rather than a dynamic, editable argument network. 
More recent AI agents, such as the AI Scientist \cite{lu2024aiscientistfullyautomated} and Scideator \cite{radensky2026humanllmcompoundscientificideation}, attempt novelty validation and even automated writing, but they require manual verification of intent alignment and offer limited support for deep logical revision \cite{10.1145/3654777.3676450}. 
\jiajun{Across these systems, explicit structure, scoped AI operations, and logic-aware review are often supported separately.}
\jiajun{This leaves limited support for continuously maintaining an author- and AI-shared representation in which argument units, their relations, and the operations applied to them remain jointly inspectable throughout drafting and revision.}
\jiajun{Our work addresses this gap through \textit{object-oriented writing}, an OOD-inspired representation and interaction paradigm instantiated in LogicTree.}
\jiajun{LogicTree combines a hierarchical object representation with structure-grounded writing assistance and structural logical review modules, allowing authors to manipulate writing objects and inspect theory-guided feedback within the same iterative workflow.}

\subsection{Structured Content Manipulation}
Structured content manipulation transforms unstructured information into modular units with defined attributes, relationships, and operations \cite{10.1145/317426.317451}. 
This paradigm is established in software engineering through \textit{object-oriented programming} \cite{booch2008object} (e.g., C++ \cite{stroustrup1986cpp}, Java \cite{gosling1996java}) and in data science environments (e.g., Jupyter notebooks \cite{kluyver2016jupyter}), where text, code, and data are organized as composable units. 
This paradigm has also influenced writing tools. 
Outlines and block-based editors shift writing from continuous linear progress to organizable units, enabling basic structural operations such as moving and deleting content \cite{engelbart1962augmenting}. 
Systems like Notion \cite{notion} extend this idea by supporting hierarchical organization and structural visualization, representing early attempts to modularize writing.

For academic writing, logical quality can be examined from both micro and macro perspectives: local coherence within specific contexts and global consistency across the entire document. 
Linguistic theory provides established frameworks for analyzing these two dimensions. 
For local coherence at the level of neighboring text segments, cohesion theory \cite{halliday2014cohesion} explains how local semantic continuity is achieved through fluent expression. 
At the level of argument structure, rhetorical structure theory (RST) \cite{rst} models text as a hierarchy of discourse relations, distinguishing between core claims and supporting information. 
The Toulmin model of argumentation \cite{toulmin2003uses} complements RST by defining the components of valid arguments: claim, grounds, warrant, backing, qualifier, and rebuttal. 
A claim is only valid when supported by grounds and licensed by a warrant. 
However, existing applications of these theories remain limited. 
VISAR \cite{10.1145/3586183.3606800} applies the Toulmin model to analyze argumentative essays, but focuses on short texts and one-time segmentation, emphasizing visualization rather than iterative verification. 
Other systems adopt structurally similar modular approaches without explicit linguistic grounding. 
Script\&Shift \cite{10.1145/3706598.3714119} and IntentFlow \cite{kim2026intentflowinvestigatingfluiddynamics} organize content around author intent, while Qin et al. \cite{10.1145/3706598.3713569} use node-graph interfaces for interactive content organization. 
These systems demonstrate the practical value of treating writing as structured manipulation of modular units, but they lack the analytical rigor that linguistic frameworks provide for assessing logical quality. 
The core challenge across both categories is that structure is treated as a static organizational scaffold rather than an active, computable property. 
Addressing this limitation, an object-oriented approach, as explored in LMCanvas \cite{kim2023lmcanvasobjectorientedinteractionpersonalize}, offers a promising direction by representing text segments as manipulable objects. 
Our work extends this direction by applying these linguistic theories to a computable review framework that both represents structure and detects logical issues, supporting iterative refinement.

\section{Formative Study}

\label{Section 3}

We conducted a formative study (N=6) to investigate the challenges researchers face during current iterative academic writing and to identify design requirements for academic writing support.
\subsection{Setup}
\textit{Recruitment \& Screening. }
We recruited participants through university forums and academic social networks. 
Candidates were required to complete a questionnaire regarding their academic background and experience with AI-assisted writing. 
We specifically selected researchers who had experience using LLMs (e.g., ChatGPT, Claude) for deep, multi-round revisions of academic manuscripts (e.g., conference papers, journal articles, or theses).

\textit{Participants. }
We recruited 6 participants (P1--P6) \jiajun{from three universities}, including 2 females and 4 males, aged 24 to 30 years.
All participants were graduate students or professional researchers in their respective domains, including Computer Science (P1--P2), Mathematics and Statistics (P3), Economics (P4), Computational Media and Arts (P5), and Medicine (P6). 
Participants reported high levels of expertise in both LLM usage (M = 4.33, SD = 1.03) and iterative document editing tools like Overleaf or Word (M = 4.5, SD = 0.84) on a 5-point Likert scale. 
This collective profile ensured that the participants possessed the necessary domain expertise to provide nuanced insights into the complex workflows of academic writing.

\textit{Interview. }
We conducted semi-structured interviews and case demonstrations lasting about 45 minutes. 
\jiajun{The study received approval from the responsible institutional ethics committee before data collection.}
\jiajun{All participants provided written informed consent for participation, audio recording, transcription, and the research use of de-identified study data.}
Participants described their academic writing workflow through concrete real-world cases, followed by questions on their use of AI-assisted tools and challenges encountered. 
Each participant received approximately US\$7 in compensation. 

\textit{Analysis. }
\jiajun{Two researchers independently performed open coding on the interview transcripts and participant case descriptions.}
\jiajun{They then compared their codes, discussed disagreements, and iteratively grouped recurring patterns into higher-level themes through inductive thematic analysis.}
\jiajun{Theme prevalence is reported below as the number of participants who endorsed a theme, rather than the frequency of coded excerpts.}
\jiajun{The resulting themes identify} challenges in aligning \jiajun{researchers' high-level intentions} with academic writing and \jiajun{directly inform} the design requirements.

\subsection{Findings}
\label{Section 3.2}

%



Participants described academic writing as an iterative process involving three tightly coupled activities: drafting, reviewing, and revising. 
They recognized that AI-assisted tools perform well in grammar correction, wording refinement, and stylistic polishing, but their support remains limited for broader tasks such as iterative organization of fragmented ideas, logic-intensive reviewing, and maintenance of cross-section consistency. 
Based on the interview results, we summarize the practices of participants at each stage and the key challenges they encountered as follows.


\textbf{C1. Mismatch between non-linear, iterative, structured drafting and linearized, text-centered AI support.}
During early drafting, most participants (5/6) started with fragmented ideas like rough outlines rather than sequential text, then iteratively organized and refined them into coherent content.
P3 stated, ``\textit{I usually start by listing some rough main points, which are often still fragmented at first. 
I then develop the content while continuously adjusting the organization and logic (of those initial points).}''

All participants (6/6) described using LLMs to explore possible logic and organization from fragmented ideas and expand partial content. 
However, such support was delivered mainly as fully verbalized text rather than explicit logical structure. 
Thus, participants had to reread lengthy outputs to reconstruct and verify the generated logic, judge whether it offered a useful direction for further development, or check whether it followed their intended writing reasoning. 
P2 noted, \textit{``I often spend more time reading and understanding what the AI generated than actually developing my own ideas.''} 
Participants also noted that LLMs often produced logical jumps or hallucinations when asked to generate long, logically connected text from rough ideas. 
Consequently, most participants (4/6) rarely relied on AI-assisted tools for organizing complete writing logic. 
As P3 summarized: \textit{``My writing process is about building a structure piece by piece, but the AI gives me a block of text that I then have to break down. It feels like we are working in opposite directions.''}

\textbf{C2: Limited support for logic-intensive reviewing and revising across local coherence and global consistency.}
After completing an initial draft, participants entered a tightly coupled review-revision process, needing to assess whether arguments were clearly conveyed, whether paragraphs supported the narrative, and whether revisions required updates elsewhere.  
Reviewing was often conducted through repeated self-reading (5/6) and feedback from collaborators (6/6). 
As P3 noted, \textit{``I usually pay particular attention to whether the overall writing framework really communicates what I want to say, and whether the sentences in each paragraph actually convey that idea.''}
Revising required authors not only to refine or restructure the problematic part of the manuscript, but also to reread large portions of the manuscript, or even the entire paper, to locate and update related statements across the document.

Participants reported that current AI-assisted tools could support this process mainly by inspecting local expressions (5/6), such as suggesting rewording, and by rewriting problematic passages (4/6) at the sentence or paragraph level. 
P5 added, \textit{``I sometimes ask it to rewrite a paragraph that feels awkward, and it often returns something that reads more smoothly.''} 
However, such support was limited for deeper logic-intensive review and revision. 
Some participants (3/6) noted that AI revisions could appear fluent without guaranteeing reasoning correctness. 
As P4 explained, \textit{``AI can rewrite a paragraph to make it sound perfect, but it might change the meaning or introduce claims that I never intended. I then have to go back and check everything carefully.''} 
Participants also observed that when revisions involved logical inconsistencies across multiple paragraphs, LLMs often performed poorly on lengthy content (5/6), could be distracted by irrelevant context (4/6), failed to identify the exact issue source (3/6), and could introduce new errors (5/6). 
This made it difficult to reliably review and revise both local coherence and global consistency with current AI support.

\subsection{Design Requirements}
\label{Section 3.3}
\jiajun{Together, \textbf{C1} and \textbf{C2} reveal a representational gap in current AI-assisted academic writing.}
\jiajun{Authors develop and evaluate an evolving logical structure, whereas interaction with AI is mediated primarily through linear prompts and generated prose.}
\jiajun{We therefore derive one foundational requirement for a shared representation and four operational requirements for drafting, reviewing, and revising through that representation.}

\jiajun{\textbf{DR1. Provide a shared, persistent, and structured intermediate representation for author-AI interaction in academic writing.}}
\jiajun{The tool should externalize the evolving structure and logic of a manuscript in a form that authors and AI can jointly reference throughout drafting, reviewing, and revision (\textbf{C1}, \textbf{C2}).}
\jiajun{This representation should persist across interactions so that authors need not repeatedly translate structural intentions into linear prompts or reconstruct the structure of AI output from prose.}

\jiajun{\textbf{DR2. Provide structured and progressive support for drafting from fragmented ideas.}}
\jiajun{The tool should accommodate authors' non-linear, iterative, and outline-centric drafting practices (\textbf{C1}).}
\jiajun{It should support progressively externalizing, organizing, expanding, and revising fragmented ideas before and alongside their realization as full prose.}

\jiajun{\textbf{DR3. Make local coherence and global consistency explicit, inspectable, and diagnosable.}}
\jiajun{The tool should help authors inspect whether arguments are clearly conveyed, whether local content supports the larger narrative, and whether a revision requires updates elsewhere in the manuscript (\textbf{C2}).}
\jiajun{It should connect potential issues to the affected manuscript parts and relations so that authors can identify where an issue occurs, what type it is, and which parts are implicated.}

\jiajun{\textbf{DR4. Provide explainable and actionable support for resolving issues in local coherence and global consistency.}}
\jiajun{Because fluent AI output may still alter intended reasoning, the tool should explain why an issue was raised and support informed revision rather than presenting a correction as authoritative (\textbf{C2}).}

\jiajun{\textbf{DR5. Provide controllable, structure-grounded AI assistance throughout the writing workflow.}}
\jiajun{The tool should ground assistance in author-recognizable writing structures, make the intended scope and proposed changes visible, and preserve authors' final judgment over whether those changes are incorporated (\textbf{C1}, \textbf{C2}).}

\section{Object-Oriented Writing}
\label{sec:oow-paradigm}

\jiajun{Motivated by the representational gap identified in our formative study and formalized in \textbf{DR1}, we propose \textit{object-oriented writing} (OOW), an OOD-inspired representation and interaction paradigm for AI-assisted academic writing (\autoref{fig:concept}).}
\jiajun{OOW introduces a shared intermediate layer through which authors and AI can refer to and operate on the evolving structure of a manuscript.}
\jiajun{Rather than requiring authors to communicate structural intentions only through linear prompts or infer the organization of AI output from prose, OOW represents a draft as a collection of identifiable and manipulable writing objects.}

\jiajun{OOW draws inspiration from object-oriented design (OOD), which represents a complex system as identifiable objects with their own state and available operations that interact through explicit relations \cite{booch2008object}.}
\jiajun{Following this representational logic, a \textbf{writing object} is a persistent and addressable unit of an academic manuscript.}
\jiajun{Its \textbf{state} records what the object currently expresses and the communicative intention or logical annotations associated with it.}
\jiajun{Its \textbf{operations} are scoped actions through which the object can be created, developed, revised, moved, inspected, or reviewed.}
\jiajun{Its \textbf{relations} connect it to other writing objects through structural, ordering, rhetorical, or argumentative dependencies.}
\jiajun{Its persistent identity allows the same unit to remain a stable referent as its state and relations change across interactions.}

\jiajun{These properties make writing objects shared referents for author-AI interaction.}
\jiajun{Authors can externalize and refine partial intentions by manipulating objects and their relations, while AI assistance can ground generation, review, and revision in the same identifiable units.}
\jiajun{Writing therefore becomes an iterative process of constructing, inspecting, and manipulating an evolving collection of related objects rather than communicating with AI solely through exchanges of linear text.}

\jiajun{Here, ``object-oriented'' denotes an object-centered representational commitment inspired by OOD, not the adoption of programming-language mechanisms such as classes, inheritance, or polymorphism.}
\jiajun{OOW does not prescribe a particular hierarchy, object granularity, relation vocabulary, interface, or AI implementation.}
\jiajun{Different systems may instantiate the paradigm through different kinds of writing objects and operations, and AI agents are not required for an implementation to follow OOW.}
\jiajun{In the next section, we present LogicTree as one concrete instantiation that uses a hierarchical object tree to support academic drafting and logical review.}

\section{\jiajun{LogicTree}}
\label{sec:logictree}

\jiajun{We instantiate OOW in LogicTree, an Overleaf-integrated academic writing prototype that operationalizes writing objects as a shared intermediate representation between authors and AI.}
\jiajun{This section first provides an overview of the system, then describes its interaction views and computational modules, and concludes with a usage scenario.}

\subsection{\jiajun{Overview}}
\label{sec:logictree-overview}

\jiajun{LogicTree coordinates the conventional Overleaf document editor (\autoref{fig:interface-screenshot} 1) with two interaction views for object-oriented authoring: an \textit{object tree} (\autoref{fig:interface-screenshot} 2) and an \textit{error list} (\autoref{fig:interface-screenshot} 3).}
\jiajun{The object tree is the primary workspace for constructing and manipulating the intermediate writing representation, while the error list presents potential logical issues and links them to the implicated objects and relations.}
\jiajun{Together, these interfaces let authors move iteratively between developing a manuscript structure and inspecting its logical quality.}

\begin{figure*}[t]
    \centering
    \includegraphics[width=0.98\textwidth]{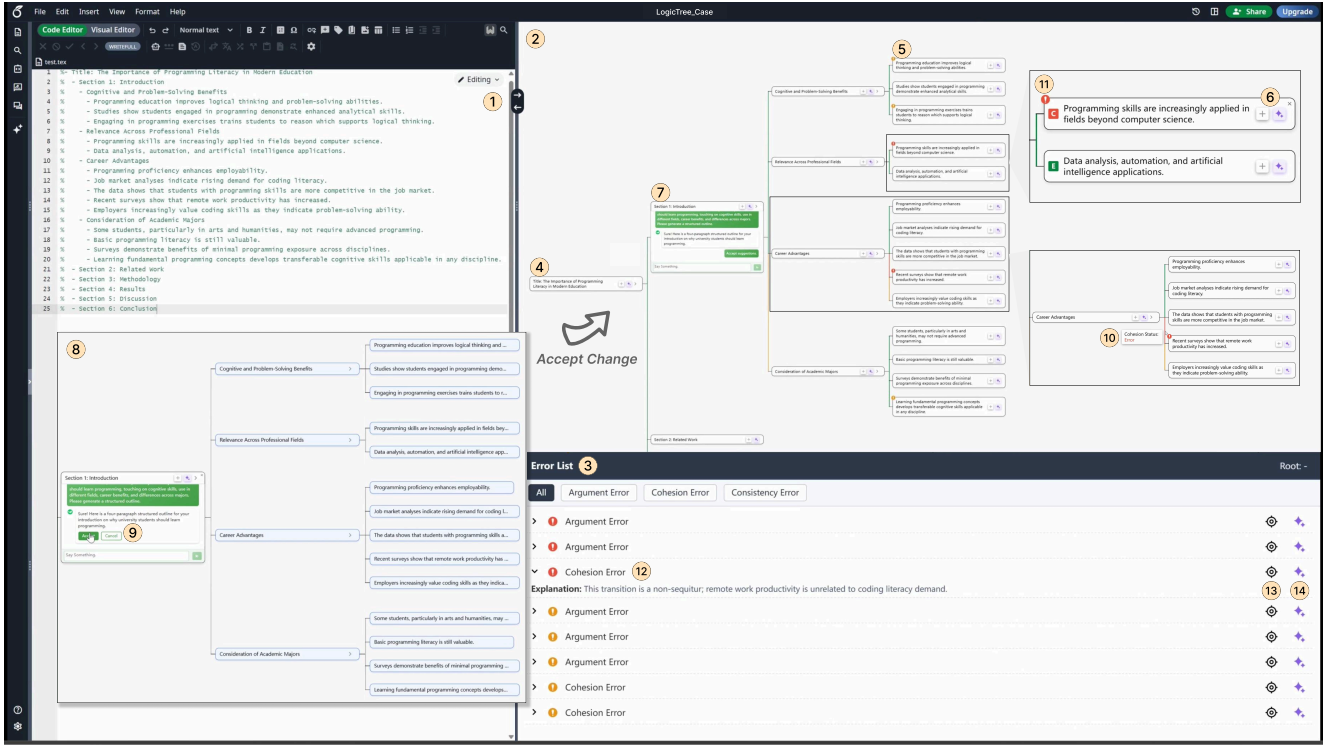}
    \caption{\jiajun{The LogicTree interface embedded in Overleaf.
    The numbered callouts indicate the (1) Overleaf Source Editor, (2) Object Tree, (3) Error List, (4) Root or Container Object, (5) Propositional Objects, (6) Object Controls, (7) Object-Scoped Conversation, (8) Blue Proposed Subtree Awaiting Acceptance, (9) Accept or Reject Proposal Controls, (10) Red Linear Cohesion Edge, (11) Argument Marker and Argumentative-Role Badges, (12) Expanded Error Entry, (13) Related-Object Locator, and (14) Issue-Grounded Revision Control.}}
    \Description{\jiajun{An annotated screenshot of LogicTree embedded in Overleaf.
    The source editor appears beside an object tree and an error list.
    The object tree shows container and propositional objects, node controls, an open conversation, a blue proposed subtree, a linear cohesion edge, and an argument marker.
    The error list shows an expanded issue entry with controls for locating the implicated objects and requesting issue-grounded revision.}}
    \label{fig:interface-screenshot}
\end{figure*}

\jiajun{LogicTree realizes each writing object as a node with a persistent identifier (\textit{Object Tree} in \autoref{sec:object-tree}), textual content, an ordered parent-child position, and a visible change state.}
\jiajun{A node may also store an argumentative role (\textit{Argument Coherence} in \autoref{sec:logical-review}), object-level conversation history, explicit relations to other nodes, and annotations produced by logical review.}
\jiajun{Higher-level \textbf{container objects} organize scopes such as sections and paragraphs (\autoref{fig:interface-screenshot} 4), while leaf-level \textbf{propositional objects} express finer-grained argument units such as Claims, Evidence, and Analysis (\autoref{fig:interface-screenshot} 5).}
\jiajun{This distinction is a design choice of LogicTree rather than a requirement of the broader OOW paradigm.}

\begin{figure*}[t]
    \centering
    \includegraphics[width=0.9\textwidth]{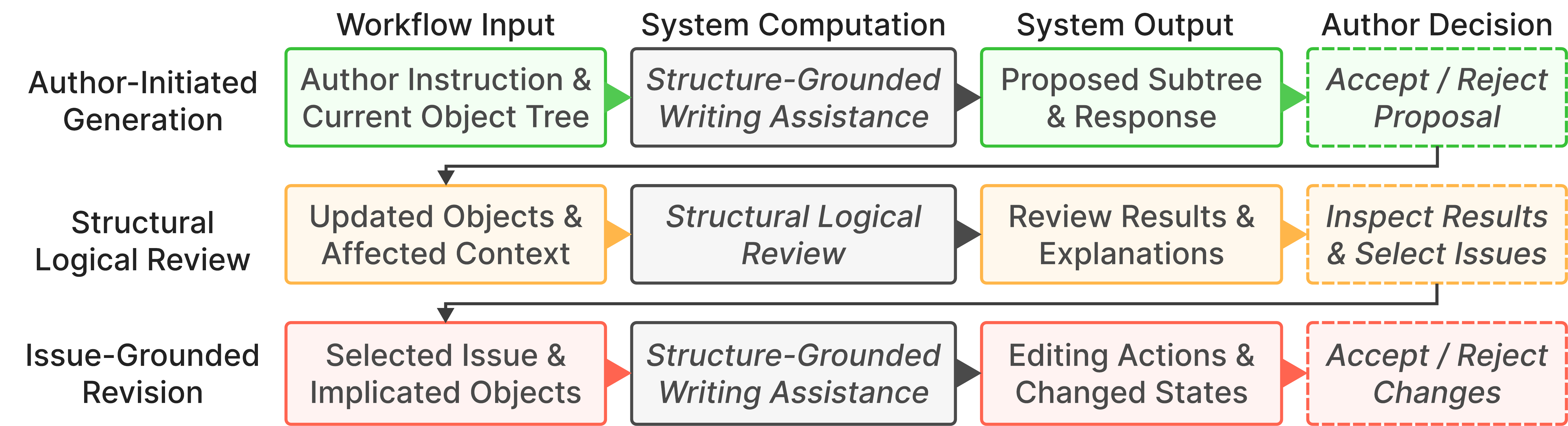}
    \caption{\jiajun{Three workflows built from two computational modules in LogicTree.
    The structure-grounded writing assistance module supports author-initiated generation and issue-grounded revision, while the structural logical review module analyzes accepted or updated objects.
    Each workflow is organized by its input, system computation, system output, and author decision.
    Solid boxes represent workflow inputs, system computations, or system outputs, while dashed boxes represent author decisions.}}
    \Description{\jiajun{A three-row workflow diagram.
    Author-initiated generation transforms an author instruction and the current object tree into a proposed subtree and response for acceptance or rejection.
    Structural logical review transforms updated objects and their affected context into review results and explanations for inspection and issue selection.
    Issue-grounded revision transforms a selected issue and its implicated objects into editing actions and changed states that the author can accept or reject.}}
    \label{fig:system-workflow}
\end{figure*}

\jiajun{The object tree and error list form the system's interaction views (\autoref{sec:interaction-views}).}
\jiajun{A structure-grounded writing assistance module and a structural logical review module operate over the same object representation (\autoref{sec:computational-modules}).}
\jiajun{\autoref{fig:system-workflow} summarizes the inputs, computations, outputs, and author decisions that connect these views and modules.}
\jiajun{This shared representation supports persistent author-AI reference (\textbf{DR1}), progressive drafting (\textbf{DR2}), inspectable and explainable review (\textbf{DR3}, \textbf{DR4}), and scoped AI assistance (\textbf{DR5}).}

\subsection{\jiajun{Interaction Views}}
\label{sec:interaction-views}

\jiajun{LogicTree provides two linked interaction views over the shared writing-object representation.}
\jiajun{The object tree supports structured authoring and inspection of object-level changes, while the error list organizes logic-review results and provides entry points for revision.}
\jiajun{Together, the two views connect manuscript construction, review, and revision without requiring authors to translate between separate representations.}

\subsubsection{\jiajun{Object Tree}}
\label{sec:object-tree}

\jiajun{Authors begin with a root object that records the manuscript's main idea (\autoref{fig:interface-screenshot} 4).}
\jiajun{They add container objects for manuscript units such as sections and paragraphs and place propositional objects beneath them to express directly writable Claims, Evidence, and Analysis.}
\jiajun{The resulting hierarchy lets authors move from an initial idea to progressively more specific content while keeping unfinished intentions visible (\textbf{DR2}).}
\jiajun{Controls attached to each object support direct content editing, child creation, subtree deletion, writing assistance, and expansion or collapse of the visible hierarchy (\autoref{fig:interface-screenshot} 6).}
\jiajun{Direct editing changes the selected object's content, whereas author-initiated generation proposes new descendants beneath it.}
\jiajun{Each object retains its identifier after content edits, which allows later assistance and review results to refer to the same writing unit (\textbf{DR1}).}

\jiajun{LogicTree can project the current hierarchy into the Overleaf source editor as nested outline comments (\autoref{fig:interface-screenshot} 1).}
\jiajun{The indentation of these comments mirrors the parent and child structure so that authors can inspect the structured draft beside the object tree.}
\jiajun{Authors continue to edit writing objects through LogicTree because arbitrary prose edits in Overleaf are not automatically parsed back into the tree.}

\jiajun{The object tree is also the entry point for the author-initiated generation workflow through its object controls (\autoref{fig:interface-screenshot} 6).}
\jiajun{Selecting an object determines the insertion point and opens the conversation retained for that object (\autoref{fig:interface-screenshot} 7).}
\jiajun{The author states how the selected object should be developed, while the current tree supplies the surrounding manuscript context.}
\jiajun{LogicTree returns the generated descendants as a blue proposed subtree beneath the selected object (\autoref{fig:interface-screenshot} 8).}
\jiajun{The proposal remains separate from accepted objects until the author accepts or rejects it through the proposal controls (\autoref{fig:interface-screenshot} 9).}
\jiajun{Accepted objects enter the established hierarchy and trigger structural logical review, while rejected objects do not affect the tree or later review results (\textbf{DR5}).}

\jiajun{Logic-review annotations are visualized on the same tree.}
\jiajun{Linear cohesion warnings and errors appear as yellow or red connections between the neighboring objects involved in the transition (\autoref{fig:interface-screenshot} 10).}
\jiajun{Argument coherence issues appear as severity markers and argumentative role badges on the implicated propositional objects (\autoref{fig:interface-screenshot} 11).}
\jiajun{Selecting a marker highlights the objects connected by the relevant argument relations and opens the corresponding explanation in the error list (\autoref{fig:interface-screenshot} 12).}
\jiajun{Global consistency issues use the error list to link an affected Claim to the other Claim with which it may conflict.}

\subsubsection{\jiajun{Error List}}
\label{sec:error-list}

\jiajun{The error list organizes potential issues in linear cohesion, argument coherence, and global consistency (\autoref{fig:interface-screenshot} 3).}
\jiajun{Each entry shows its issue category and severity in the collapsed state.}
\jiajun{Expanding an entry reveals the implicated object or pair and a concise explanation of the detected transition, support gap, or inconsistency (\autoref{fig:interface-screenshot} 12).}
\jiajun{The related-object locator expands the paths to the implicated objects in the object tree so that the author can inspect the issue in its structural context (\autoref{fig:interface-screenshot} 13).}
\jiajun{Conversely, selecting a tree marker opens the corresponding error entry.}
\jiajun{This bidirectional link connects each diagnosis to the structural context from which it was produced (\textbf{DR3}, \textbf{DR4}).}

\jiajun{Authors can respond to an issue by revising the implicated objects manually or selecting the issue-grounded revision control (\autoref{fig:interface-screenshot} 14).}
\jiajun{For supported linear cohesion and argument coherence issues, the system passes the implicated objects, their local structural context, and the diagnostic explanation to the structure-grounded writing assistance module.}
\jiajun{The control returns a contextualized revision and maps it to additions, modifications, or removals shown as pending change states in the object tree.}
\jiajun{The author can accept, ignore, or further revise the proposed change while inspecting it together with the original issue.}

\subsection{\jiajun{Computational Modules}}
\label{sec:computational-modules}

\jiajun{As shown in \autoref{fig:system-workflow}, LogicTree organizes its operation into three workflows that reuse two computational modules.}
\jiajun{Each module combines an LLM-based agent with deterministic procedures for context construction, output validation, and object-state management.}
\jiajun{The structure-grounded writing agent and logic-review agent perform model-based generation or diagnosis, while their enclosing modules coordinate the complete system workflows.}
\jiajun{The structure-grounded writing assistance module supports both author-initiated generation and issue-grounded revision.}
\jiajun{Author-initiated generation transforms an author instruction and the selected object context into a proposed subtree.}
\jiajun{Issue-grounded revision transforms a selected review issue and its implicated objects into proposed editing actions.}
\jiajun{The structural logical review module forms the third workflow and analyzes accepted or directly updated objects for local coherence and global consistency.}
\jiajun{Each workflow follows the same four stages of input, computation, output, and author decision.}
\jiajun{The modules return object-level proposals and annotations that remain inspectable before authors accept changes or initiate further revision.}

\subsubsection{\jiajun{Structure-Grounded Writing Assistance}}
\label{sec:outline-editing} 
\jiajun{LogicTree supports two structure-grounded writing workflows, namely author-initiated generation and issue-grounded revision.}

\paragraph{Author-initiated generation} \jiajun{Author-initiated generation begins when an author wants to expand a selected object with additional manuscript content.}
\jiajun{The author invokes writing assistance on that object and describes the content to develop.}
\jiajun{LogicTree uses the selected object as the insertion scope and returns a proposed subtree beneath it.}

\jiajun{The selected object identifier fixes the insertion position.}
\jiajun{LogicTree combines the author's instruction and the conversation stored for that object with the identifiers, textual content, and hierarchical relations between parent and child objects in the current tree.}
\jiajun{This input lets the structure-grounded writing agent interpret the local request in relation to the surrounding manuscript structure.}
\jiajun{The generation prompt asks the writing agent to begin with immediate children of the selected object, develop them into progressively more specific descendants, and end in propositional objects containing directly writable key sentences.}
\jiajun{It constrains the generated subtree to at most three levels beneath the selected object with multiple siblings at each developed level.}

\jiajun{The writing agent returns a JSON object containing a short response and a recursively nested list of proposed child objects.}
\jiajun{LogicTree parses this output and accepts only recursively nested child records with nonempty textual content.}
\jiajun{The system then assigns each proposed object a persistent identifier and an added change state.}
\jiajun{A separate classification call assigns provisional argumentative roles to generated leaf objects by considering their common parent and sibling content together.}
\jiajun{The resulting subtree appears in blue beneath the selected object so that the author can accept or reject it before it becomes part of the established tree.}

\jiajun{Acceptance changes the proposed objects to the normal state and triggers relation construction and structural logical review.}
\jiajun{This sequence keeps the selected object as the scope defined by the author while using the remaining tree only as context (\textbf{DR5}).}
\jiajun{It also prevents rejected generations from affecting later review results.}

\paragraph{Issue-grounded revision.} \jiajun{Assistance initiated from an issue begins with a specific warning or error rather than an open author instruction.}
\jiajun{For supported linear cohesion and argument coherence issues, LogicTree provides the structure-grounded writing assistance module with the implicated object identifiers and content, their parent and sibling context, the severity, and the diagnostic explanation.}
\jiajun{Within the module, the writing agent first produces an actionable repair suggestion and then converts it into one or more structured add, modify, or remove actions.}
\jiajun{For example, when a Claim has Evidence but no Analysis, the writing agent can propose an Analysis object that explains why the Evidence supports the Claim.}
\jiajun{LogicTree maps every action to explicit object identifiers and displays its change state in the object tree for author inspection before incorporation.}

\subsubsection{\jiajun{Structural Logical Review}}
\label{sec:logical-review}

\begin{figure*}[tbh]
    \centering
    \includegraphics[width=0.95\textwidth]{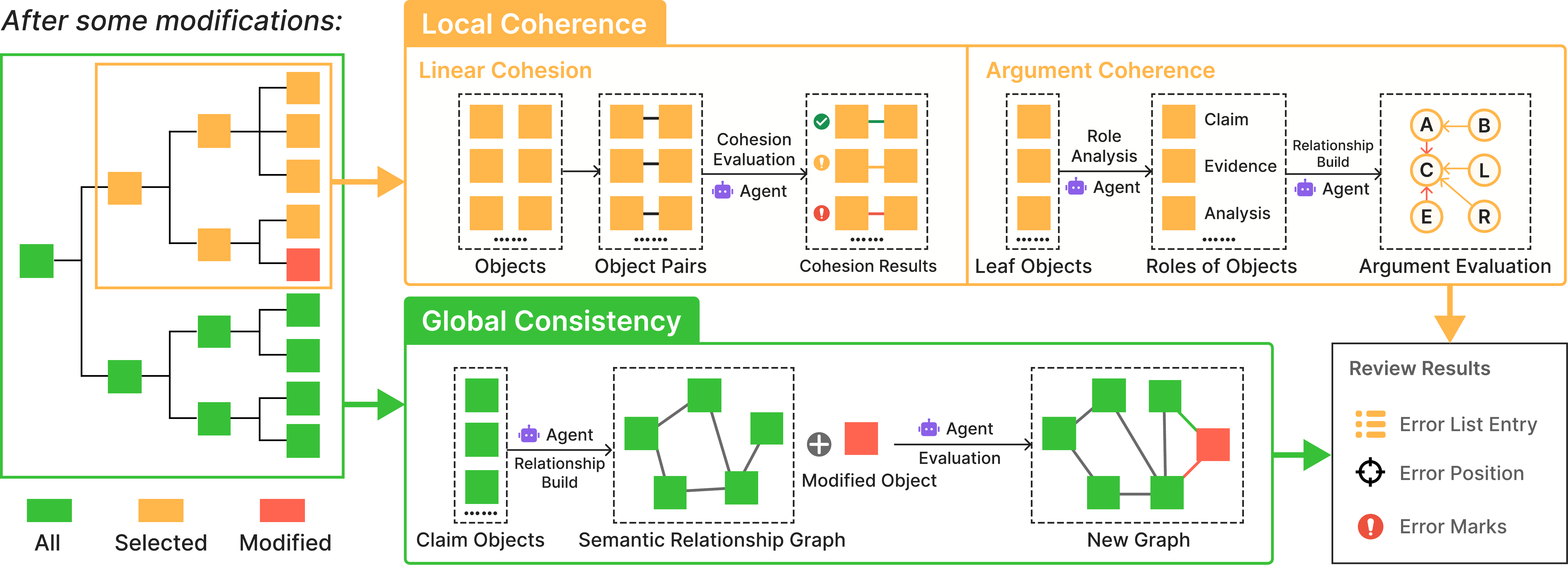}
    \caption{\jiajun{Structural logical review in LogicTree.
    Local review evaluates transitions between adjacent subtrees and argument structures among propositional objects in the affected context.
    Global review compares affected Claims with the other Claims in the current object tree.}}
    \Description{\jiajun{A workflow diagram showing how LogicTree checks linear cohesion, argument coherence, and global consistency over the writing-object tree.}}
    \label{fig:logic_review}
\end{figure*}

\jiajun{Drawing on theories of textual coherence and discourse relations \cite{hasan1984coherence,rst}, the structural logical review module examines accepted generated objects and directly updated leaf objects through local coherence and global consistency (\autoref{fig:logic_review}).}
\jiajun{The updated objects form the affected set that determines which checks are recomputed.}
\jiajun{Local review inspects transitions and argument structures in the affected structural context.}
\jiajun{Global review compares each affected Claim with Claim objects elsewhere in the current tree.}
\jiajun{Every review routine returns object identifiers, a severity label, and a concise explanation.}
\jiajun{LogicTree validates the returned identifiers and the eligible comparison scope before storing warning or error annotations on the corresponding objects or relations.}
\jiajun{An object or relation without such an annotation received no reported issue under the applied check.}

\paragraph{\jiajun{Linear Cohesion.}}
\jiajun{Following cohesion theory \cite{halliday2014cohesion}, linear cohesion evaluates whether adjacent manuscript units form a comprehensible semantic progression.}
\jiajun{LogicTree constructs candidate pairs from adjacent sibling subtrees and from the transition between adjacent parent subtrees within the modified parent.}
\jiajun{The logic review prompt receives the identifiers, content, and descendant hierarchy of this scope and asks the logic-review agent to return only insufficiently coherent candidate pairs.}
\jiajun{A transition is a warning when the two units retain a plausible topical or argumentative connection but that connection is implicit, abrupt, or insufficiently developed.}
\jiajun{A transition is an error when the second unit does not provide a plausible continuation of the first or introduces content unrelated to the surrounding topic.}
\jiajun{For example, moving from evidence about demand for programming skills to an isolated statement about remote work productivity is classified as an error because the latter does not continue the former topic.}
\jiajun{For every returned pair, LogicTree verifies that both identifiers exist, that the pair is adjacent, and that at least one side contains an affected object.}
\jiajun{A validated warning or error is rendered as a colored connection between the two objects, and the same issue appears in the error list with its explanation.}

\paragraph{\jiajun{Argument Coherence.}}
\jiajun{LogicTree operationalizes argument coherence through discourse relations \cite{rst} and six roles adapted from the Toulmin model \cite{toulmin2003uses}.}
\jiajun{A Claim is the debatable assertion that the author asks the reader to accept.}
\jiajun{Evidence consists of independently verifiable facts, observations, data, or examples.}
\jiajun{Analysis states the reasoning that explains how or why Evidence supports a Claim.}
\jiajun{Backing supplies an authoritative source, theory, or principle for that reasoning.}
\jiajun{A Limitation restricts the scope or strength of a Claim, while a Rebuttal identifies an exception or competing position.}

\jiajun{The module groups leaf objects by their direct parent and asks the logic-review agent to assign one role to each object using the parent topic and sibling content as context.}
\jiajun{The agent then proposes semantically justified relations that conform to the five role relation patterns in \autoref{table:edges}.}
\jiajun{Evidence supports a Claim, Analysis licenses a Claim, and Backing supports Analysis.}
\jiajun{Rebuttal and Limitation objects connect to a Claim through the stored attack relation, where Rebuttal opposes the Claim and Limitation constrains its applicability.}
\jiajun{The first three relation types are constructed among leaf siblings under the same parent.}
\jiajun{Rebuttal and Limitation relations may connect to an affected Claim within the same section.}
\jiajun{LogicTree accepts a proposed relation only when its endpoints are leaf objects, its direction matches their roles, and its structural scope satisfies these constraints.}

\jiajun{The system applies deterministic criteria after role and relation construction.}
\jiajun{It reports an error when a Claim has no incoming support from Evidence, when a Claim has no incoming license from Analysis, or when a propositional object other than a Claim has no relation to another object.}
\jiajun{It reports a warning when an Analysis has no incoming support from Backing or when a linguistically strong Claim has no Limitation.}
\jiajun{Strong Claims are identified through categorical or necessity expressions such as \textit{must}, \textit{all}, \textit{necessary}, and \textit{essential}.}
\jiajun{For example, the Claim ``Programming literacy is essential for all students'' receives an error if it has Evidence but no Analysis and receives a warning if it has no Limitation.}
\jiajun{Objects with detected issues receive severity markers and role badges in the object tree, while selecting a marker highlights the related objects and opens the explanation in the error list.}
\jiajun{These checks operationalize the presence and arrangement of argumentative support rather than the truth of the Claim.}

\begin{table}[h]
\centering
\caption{\jiajun{Five role-relation patterns used for argument-coherence checks.}}
\begin{tabular}{p{0.3\columnwidth} p{0.3\columnwidth} p{0.15\columnwidth}}
\toprule
\textbf{Source} & \textbf{Target} & \textbf{Type} \\
\midrule
Evidence   & Claim    & Support \\
Analysis   & Claim    & License \\
Backing    & Analysis & Support \\
Rebuttal   & Claim    & Attack \\
Limitation & Claim    & \jiajun{Attack} \\
\bottomrule
\end{tabular}
\label{table:edges}
\end{table}

\paragraph{\jiajun{Global Consistency.}}
\jiajun{The Toulmin model treats a Claim as the proposition advanced for acceptance and relates other argumentative roles to that proposition \cite{toulmin2003uses}.}
\jiajun{Global consistency therefore evaluates whether an affected Claim can coexist with Claims elsewhere in the manuscript.}
\jiajun{LogicTree limits this comparison to Claims because they express the manuscript's substantive positions, while the relations of Evidence, Analysis, Backing, Rebuttal, and Limitation to Claims are assessed during argument coherence review.}
\jiajun{LogicTree forms the comparison pool from all leaf objects currently classified as Claims and pairs each affected Claim with every other Claim in that pool.}
\jiajun{LogicTree represents these candidate Claim pairs in a semantic relationship graph and links each affected Claim to that graph for evaluation.}
\jiajun{The logic-review agent receives the identifiers and content of these Claims and returns only conflicting or potentially conflicting pairs.}
\jiajun{A direct contradiction is recorded as an error.}
\jiajun{A likely tension, scope mismatch, or overgeneralization is recorded as a warning.}
\jiajun{For example, an unrestricted Claim that programming is necessary for every student may receive a warning when another Claim states that advanced programming is unnecessary for students in some disciplines.}
\jiajun{LogicTree verifies that the first identifier belongs to the affected Claim set, that both endpoints are Claim leaves, and that the two identifiers differ before storing the issue and explanation.}
\jiajun{The graph organizes candidate comparisons and does not impose a formal logical relation between the Claims.}
\jiajun{The module uses the logic-review agent for diagnosis rather than formal logical inference.}

\subsection{\jiajun{Usage Scenario}}
\label{sec:usage-scenario}

\begin{figure*}[tbh]
    \centering
    \includegraphics[width=0.95\textwidth]{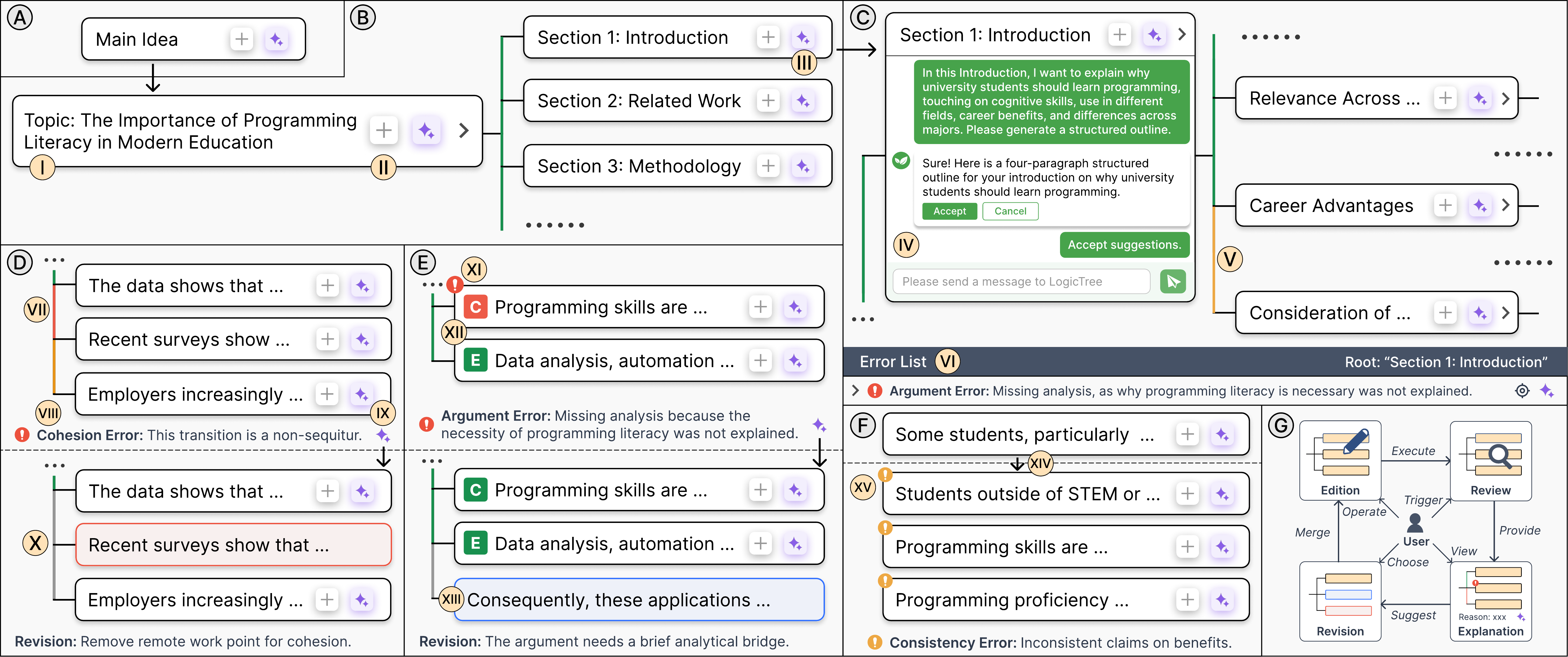}
    \caption{User interaction flow with \jiajun{LogicTree}.
    The user starts from a root \jiajun{container object} (A) and builds the initial \jiajun{hierarchy} (B).
    After accepting \jiajun{writing assistance proposals}, the \jiajun{structural logical review module} highlights issues in the \jiajun{object} tree and error list (C).
    The user identifies and resolves three types of \jiajun{potential issues}: linear cohesion \jiajun{issues} (D), argument coherence \jiajun{issues} (E) and global consistency \jiajun{issues} (F).
    Through iterative cycles of editing, reviewing, and revising (G), the user completes the academic writing task.
    }
    \Description{A schematic of the \jiajun{LogicTree} user interface and its object-grounded writing and review workflow, with labeled elements A through G.}
    \label{fig:scenario_interface}
\end{figure*}

\jiajun{Frank, a doctoral student in education, opens LogicTree with the main idea that programming literacy is increasingly important in modern education.}
\jiajun{He edits the root object and adds section-level container objects to establish an initial manuscript structure (\autoref{fig:scenario_interface} A and B).}
\jiajun{Because his Introduction remains incomplete, he invokes writing assistance on that object and describes the ideas he wants to develop.}
\jiajun{LogicTree presents proposed paragraph and propositional objects beneath the selected object, and Frank inspects them before acceptance (\autoref{fig:scenario_interface} C).}

\jiajun{After the accepted objects are reviewed, Frank notices a red linear cohesion edge connecting a statement about programming skills in the job market with an unrelated statement about remote-work productivity (\autoref{fig:scenario_interface} D).}
\jiajun{He selects the edge to read the explanation in the error list and requests repair assistance.}
\jiajun{The structure-grounded writing assistance module proposes removing the off-topic object, which Frank reviews before accepting.}

\jiajun{Frank then inspects an argument marker showing that a Claim about the broader value of programming literacy has Evidence but lacks Analysis (\autoref{fig:scenario_interface} E).}
\jiajun{The linked error entry explains the missing connection, and the structure-grounded writing assistance module proposes an Analysis object that Frank accepts after inspection.}

\jiajun{Later, Frank strengthens a Claim about students outside technical fields.}
\jiajun{The consistency review flags tensions between this revision and Claims elsewhere in the manuscript (\autoref{fig:scenario_interface} F).}
\jiajun{Frank examines the implicated objects and revises the statement himself.}
\jiajun{Through repeated cycles of construction, inspection, and revision, he uses the same object representation to coordinate his intentions with AI assistance while retaining final control over the manuscript (\autoref{fig:scenario_interface} G).}

\section{Evaluation}

We evaluate the \jiajun{integrated LogicTree workflow} through a user study with 12 participants (P1--P12) from diverse academic backgrounds.
Specifically, we investigate the following research questions:

\begin{itemize}[leftmargin=*, labelindent=0pt, itemindent=0pt]
    \item \jiajun{How accurately does the structural logical review module annotate argumentative roles, relations, and logical issues?}
    \item How does \jiajun{LogicTree} support users in organizing and revising academic text according to their writing intentions, and how do users perceive its usability?
    \item How does the logical quality of academic texts produced with \jiajun{LogicTree} compare with texts produced using a conversational baseline \jiajun{in the controlled study tasks}?
\end{itemize}

\subsection{Experiment Settings}

We conducted a within-subjects study with 12 participants to compare the \jiajun{complete LogicTree workflow} against a conversational baseline.
Participants completed two academic writing tasks, after which we analyzed their performance and subjective feedback.
Our use of OpenAI Codex for prototype implementation and post-study data aggregation and visualization is detailed in the Generative AI Use Disclosure at the end of the paper.

\textbf{Participants.} 
We recruited 12 participants (P1--P12; 8 males, 4 females) through social media and word‑of‑mouth. 
They were all graduate students and researchers with diverse backgrounds, such as computer science, medicine, optical engineering, and marketing. 
According to their self‑ratings on a 7‑point Likert scale (1: lowest extent, 7: greatest extent), participants were experienced with academic paper writing (M = 5.42, SD = 1.16) and had prior experience using LLMs to assist with academic writing (M = 5.58, SD = 1.00). 
All participants had authored at least one peer‑reviewed English paper and were familiar with Overleaf, ensuring they met the proficiency requirements for the academic writing tasks.

\textbf{Baseline and Apparatus.} 
We adopted a \jiajun{within-subjects} design with two conditions.
In the \jiajun{LogicTree} condition, participants used our prototype \jiajun{integrated into Overleaf}.
In the Baseline condition, participants used a system functionally equivalent to conversational AI tools like ChatGPT for academic writing. 
\jiajun{The Baseline and LogicTree conditions used the same foundation model (i.e., qwen3-next-80b-a3b-instruct), and the baseline preserved conversation context and could access the same task materials.}
\jiajun{The comparison therefore evaluates LogicTree as an integrated workflow rather than isolating the causal contribution of each feature.}

\textbf{Tasks.} 
To reduce the influence of domain‑specific knowledge and ensure controllable experimental conditions, the evaluation comprised two academic writing tasks. 
For each task, participants received an unordered information package containing ten snippets drawn from top‑tier academic journals, government documents and authoritative media. 
Task A focused on the reliability and interpretability of medical AI, and Task B on the ethical dilemmas of autonomous driving. 
Participants were instructed to write \jiajun{an} academic discussion section of 500--800 words, use only the provided snippets for empirical content, and aim for the highest possible logical quality.
They were not required to use every snippet.

\subsection{Procedure}
\jiajun{The evaluation received approval from the responsible institutional ethics committee before data collection.}
\jiajun{All participants provided written informed consent for participation, audio and screen recording, and the research use of de-identified study data.}
We employed a counterbalanced \jiajun{within-subjects} design, randomly assigning participants to one of four sequences balancing tool (\jiajun{LogicTree} or Baseline) and task (A or B) order.
Prior to the experimental procedures, participants took part in a 10‑minute training session that introduced the two systems and included a brief practice task.
Before each timed task, participants reviewed its information package and could ask clarifying questions.
They began the 20-minute writing period once they indicated that they had no further questions, after approximately 5 minutes of preparation on average.
Following the training, participants worked on the first task for 20 minutes using the assigned system, after which they immediately completed \jiajun{post-task} questionnaires including SUS~\cite{brooke2013sus}, \jiajun{NASA-TLX}~\cite{hart2006nasa}, and a contextualized usability questionnaire, with an additional feature questionnaire for the \jiajun{LogicTree} condition.
They then proceeded to the second task for 20 minutes using the other system and completed the same questionnaires. 
The evaluation concluded with a 10–15 minute semi‑structured interview, which was audio‑recorded. 
The interview focused on gathering overall impressions of participants from each system, the strategies they used for constructing logical arguments, and the suggestions they provided for improvement. 
The entire experimental procedure lasted approximately 75 minutes. 
Participants were compensated with around US\$10/hour. 
All task sessions were screen‑recorded.

We then evaluated the academic texts produced by participants using three LLMs (i.e., deepseek‑v3.2, gpt‑5.4, and glm‑5), each scoring the texts across multiple logical dimensions. 
To reduce the impact of stochastic variability, each model scored every text for 100 independent rounds. 
\jiajun{We also evaluated the structural logical review module on the 12 texts produced in the LogicTree condition.}
\jiajun{Two researchers created a consensus human reference, and we compared three runs of the module with this reference using agreement and classification metrics.}
\jiajun{The annotation procedure, metrics, and results are reported in \autoref{sec:logic-review-accuracy}.}
The analysis combined questionnaire responses, interview transcripts, model ratings of logical quality, and recorded interaction behavior. 
Detailed questionnaires and scoring dimensions are provided in the supplementary materials.

\subsection{Results}

We report the evaluation results including \jiajun{accuracy of the logical review}, subjective ratings, logical quality of academic texts and participant feedback.

\begin{table*}[tp]
\centering
\normalsize
\caption{\jiajun{Human annotation agreement and structural logical review module performance across four annotation components.
The table reports sample size, percent agreement, Cohen's $\kappa$, Gwet's AC1, system accuracy, and F1.
Agreement, accuracy, and F1 are percentages.
System results are reported as mean (SD) over three runs.}}
\label{tab:logic_review_accuracy}
\setlength{\tabcolsep}{5pt}
\renewcommand{\arraystretch}{1.1}
\begin{tabular}{lcccccc}
\toprule
& & \multicolumn{3}{c}{\textbf{Human annotation}} & \multicolumn{2}{c}{\textbf{System performance}} \\
\cmidrule(lr){3-5}\cmidrule(lr){6-7}
\textbf{Component} & \textbf{$N$} & \textbf{Agreement} & \textbf{Cohen's $\kappa$} & \textbf{AC1} & \textbf{Accuracy} & \textbf{(Macro-)F1} \\
\midrule
Role classification        & 239 & 89.54 & 0.845 & 0.879 & 92.05 (0.42) & 85.90 (2.92) \\
Relation construction      & 205 & 85.37 & 0.771 & 0.836 & 77.56 (0.49) & 82.41 (1.26) \\
Linear cohesion            & 228 & 88.16 & 0.530 & 0.865 & 85.53 (0.44) & 75.97 (1.17) \\
Global consistency         & 170 & 97.06 & 0.750 & 0.969 & 93.53 (0.59) & 56.78 (1.35) \\
\midrule
Structural review (pooled) & 603 & N/A   & N/A   & N/A   & 85.07 (0.50) & N/A \\
\bottomrule
\end{tabular}
\normalsize
\end{table*}

\subsubsection{\jiajun{Accuracy of the logical review}}
\label{sec:logic-review-accuracy}

\jiajun{We evaluated the reliability of the structural logical review module by first establishing a consensus human reference and then comparing the module's outputs with that reference.
The results are shown in \autoref{tab:logic_review_accuracy}.
}

\jiajun{The annotation set covered the 12 texts produced in the LogicTree condition and used a task-specific unit for each review component.}
\jiajun{It comprised 239 propositional objects for role classification, 205 candidate object pairs for relation construction, 228 boundaries between adjacent objects or subtrees for linear cohesion, and 170 Claim pairs for global consistency.}
\jiajun{Two researchers independently labeled these units without inspecting the system outputs.}
\jiajun{They then resolved disagreements through discussion to form the consensus reference used for system evaluation.}
\yanna{Based on their independent annotations before disagreement resolution, we assessed inter-annotator agreement using three complementary measures: percent agreement, Cohen's $\kappa$, and Gwet's AC1.
Percent agreement measures the proportion of instances for which the two annotators assigned the same label.
Cohen's $\kappa$ further accounts for agreement that could occur by chance given the annotators' label distributions.
However, $\kappa$ can be sensitive to highly imbalanced label distributions, as in the linear cohesion and global consistency annotation sets where ``No Issue'' was the majority label.
We therefore additionally report Gwet's AC1, a chance-corrected agreement measure that is more robust to such label imbalance. 
For both Cohen's $\kappa$ and Gwet's AC1, values closer to 1 indicate stronger agreement.}
\jiajun{As shown in \autoref{tab:logic_review_accuracy}, percent agreement ranged from 85.37\% to 97.06\%, and AC1 ranged from 0.836 to 0.969 across the four tasks.}
\jiajun{Linear cohesion had a lower Cohen's $\kappa$ of 0.530 but retained 88.16\% agreement and an AC1 of 0.865, a pattern consistent with the prevalence of ``No Issue'' labels.}

\jiajun{We ran the frozen structural logical review module three times with the same model, prompts, and parameters and evaluated every prediction against the consensus human annotation using anonymous object or pair identifiers.}
\yanna{We evaluated performance using accuracy and F1 score and report the mean and standard deviation across runs.
Accuracy measures the proportion of predictions that exactly match the consensus human reference.
F1 jointly considers precision and recall and is therefore more informative when the labels are imbalanced. 
For role classification, which contains multiple argumentative roles, we report macro-F1 by computing the F1 score for each role separately and then averaging across roles, so that each role contributes equally regardless of its frequency.
For relation construction, linear cohesion, and global consistency, we report F1 for detecting the presence of a relation or issue, treating ``No Relation'' or ``No Issue'' as the negative class.
}
\jiajun{Pooling the 603 relation and issue decisions from relation construction, linear cohesion, and global consistency yielded an accuracy of 85.07\% with an SD of 0.50 percentage points (\autoref{tab:logic_review_accuracy}).}
\jiajun{Role classification reached 92.05\% accuracy and 85.90\% macro-F1, showing that the module usually recovered the argumentative role assigned in the human reference.}
\jiajun{Exact relation construction was the most demanding component and reached 77.56\% accuracy because both relation type and direction had to match.}
\jiajun{Global consistency reached 93.53\% accuracy but 56.78\% issue F1 because only 11 of 170 Claim pairs contained a Warning or Error.}
\jiajun{This result indicates that the module usually recognized ``No Issue'' pairs but had greater difficulty identifying the less frequent consistency issues.}
\jiajun{Overall, the remaining errors were concentrated in exact relation matching and rare consistency issues rather than distributed uniformly across the review pipeline.}

\begin{figure*}[t]
    \centering
    \includegraphics[width=\textwidth]{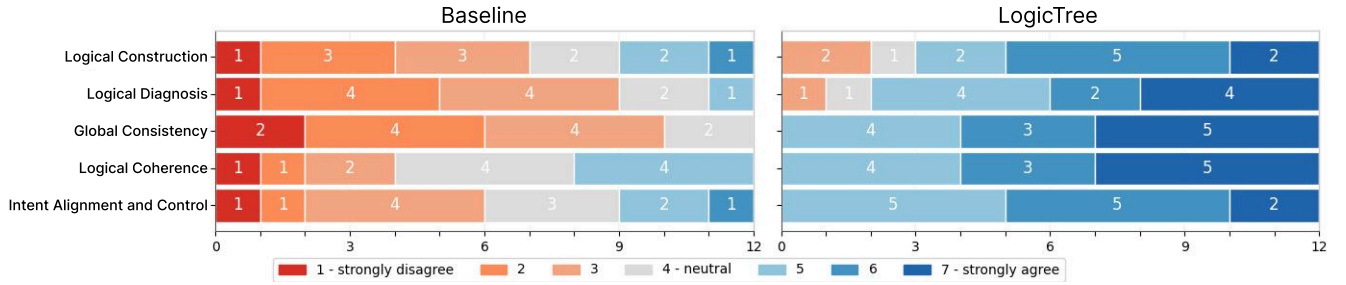}
    \caption{User ratings of contextualized usability for Baseline and \jiajun{LogicTree} on a 7-point Likert scale.
    Each participant's mean score for a dimension was rounded to the nearest Likert category for the stacked distributions.}
    \Description{User ratings on Baseline and \jiajun{LogicTree}.}
    \label{fig:evaluation_compare}
\end{figure*}

\begin{figure}[t]
    \centering
    \includegraphics[width=\linewidth]{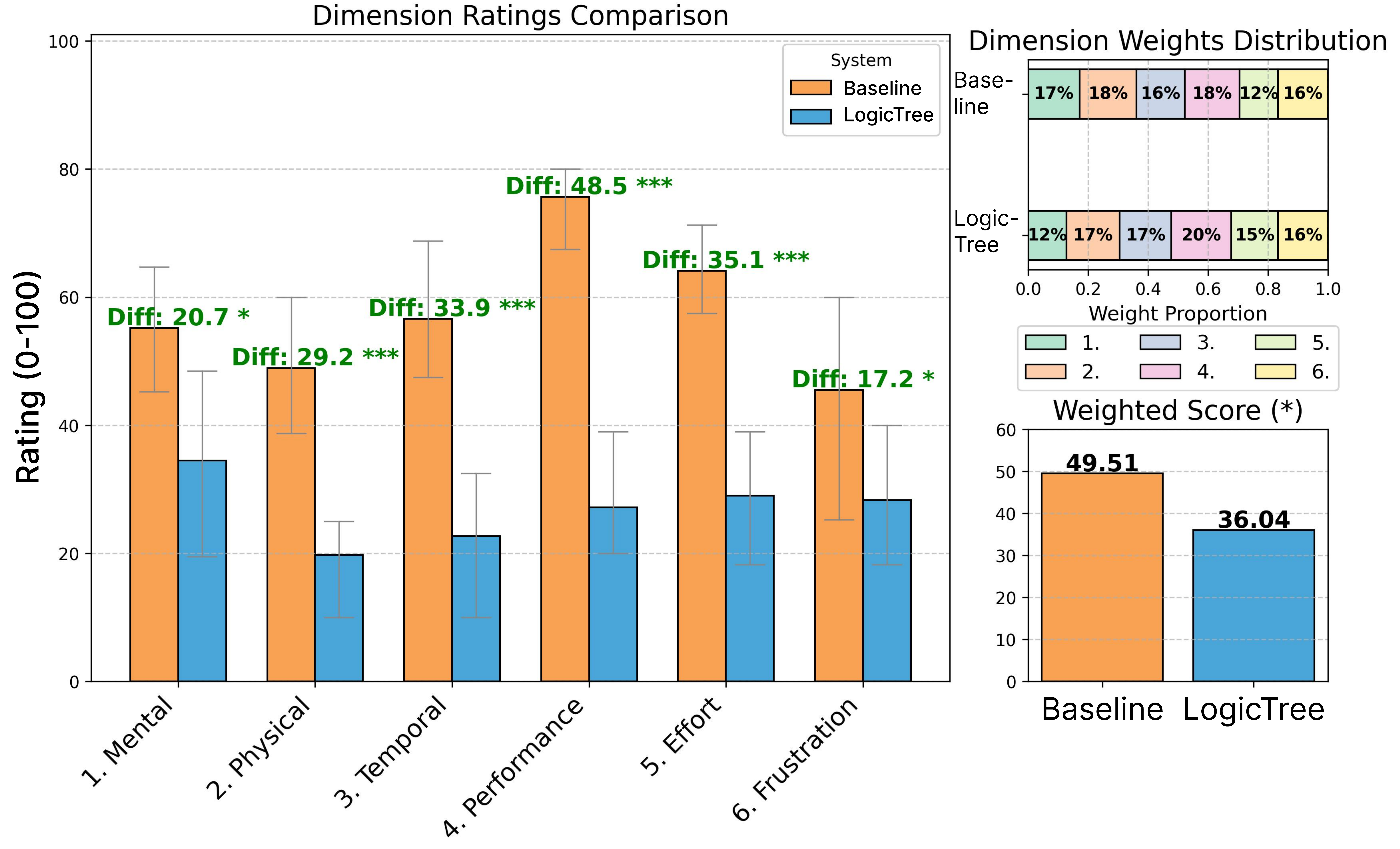}
    \caption{User ratings in NASA-TLX.}
    \Description{Rating results in NASA-TLX.}
    \label{fig:evaluation_nasa}
\end{figure}

\begin{figure}[t]
    \centering
    \includegraphics[width=\linewidth]{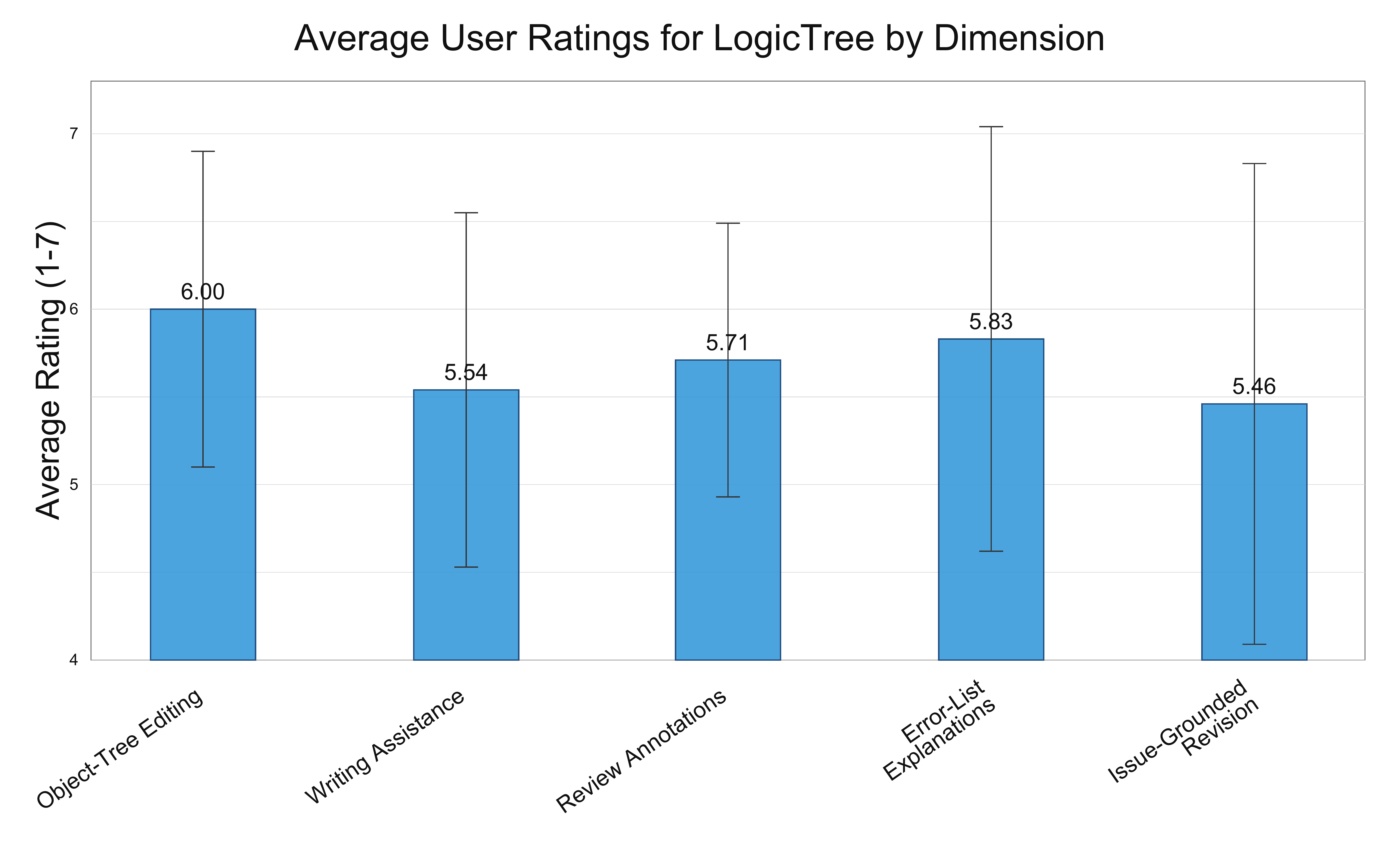}
    \caption{User ratings of five \jiajun{LogicTree} features on a 7-point Likert scale.
    }
    \Description{A bar chart reports mean ratings on a seven-point scale for five LogicTree features.
    Object-Tree Editing received 6.00, Writing Assistance 5.54, Review Annotations 5.71, Error-List Explanations 5.83, and Issue-Grounded Revision 5.46.
    Vertical error bars show the standard deviation of participant ratings.}
    \label{fig:evaluation_function}
\end{figure}

\subsubsection{Subjective ratings}
We first assessed general system usability using the System Usability Scale (SUS)~\cite{brooke2013sus}. 
The results showed that \jiajun{LogicTree} received a higher SUS score (M = 75.21, SD = 16.06) than Baseline (M = 58.54, SD = 21.83), placing \jiajun{LogicTree} in the range of good usability~\cite{sauro2016quantifying}.
To further compare system usability and effectiveness aligned with research questions and design requirements, we developed a contextualized usability questionnaire with 12 questions across five dimensions, 2 to 3 per dimension. 
For each dimension, we calculated the score as the average of its constituent questions. 
We then conducted Shapiro-Wilk tests to assess normality (all $p > 0.05$) and used paired $t$-tests ($\alpha = 0.05$). 
\jiajun{LogicTree} received significantly higher ratings than Baseline ($p < 0.05$) across all five dimensions (\autoref{fig:evaluation_compare}).
We also assessed task workload using the NASA-TLX questionnaire \cite{hart1988development}, which measures six dimensions of task load. 
Following the same statistical procedure, the results indicated significantly lower \jiajun{reported} workload with \jiajun{LogicTree} than with Baseline ($p < 0.05$) for all six dimensions (\autoref{fig:evaluation_nasa}).
\jiajun{To evaluate the perceived usefulness of individual LogicTree features, we designed a 7-point Likert questionnaire with two items for each of five features, namely object-tree editing, structure-grounded writing assistance, object-tree review annotations, error-list explanations, and issue-grounded revision.}
Each feature score was calculated as the average of its two questions. 
Users responded positively (M = 5.46--6.00, SD = 0.78--1.37) to the \jiajun{LogicTree} features (\autoref{fig:evaluation_function}).
\jiajun{These ratings describe perceived usefulness but do not isolate the causal effect of individual features.}

\subsubsection{Logical quality of academic texts}
To evaluate the logical quality of academic texts produced by participants using \jiajun{LogicTree} and Baseline, we employed three large language models (i.e., deepseek-v3.2, gpt-5.4, and glm-5).
Each model independently rated every text in 100 rounds using a seven-item, 7-point Likert questionnaire, with two items each assessing Linear Cohesion, Argument Coherence, and Global Consistency and one item assessing Overall Logical Quality.
Scores for each dimension were calculated by averaging its corresponding items.
Across the four dimensions, texts produced in the \jiajun{LogicTree} condition received higher average scores than those in the Baseline condition from all three models (deepseek-v3.2: 6.03--6.27 vs. 3.49--4.24; gpt-5.4: 6.56--6.75 vs. 2.57--3.32; glm-5: 5.82--6.08 vs. 3.53--4.31; \autoref{tab:results}).
The intra-class correlation coefficients (ICC) ranged from 0.961 to 0.998, indicating \jiajun{stability of the model-generated scores across repeated evaluation rounds}\jiajun{; these values do not by themselves establish the validity of the ratings}.

\begin{table*}[tp]
\centering
\normalsize

\caption{The average scores (Avg.) and intra-class correlation coefficients (ICC) for logical quality across four dimensions generated by three large language models (deepseek-v3.2, gpt-5.4, and glm-5) for academic texts produced in \jiajun{LogicTree} and Baseline conditions.
Standard deviations (in parentheses) are calculated from the scores aggregated across 100 rounds.
\jiajun{ICC reflects score stability rather than rating validity.}}
\label{tab:results}
\normalsize
\setlength{\tabcolsep}{3pt} 
\renewcommand{\arraystretch}{1.1} 
\begin{tabular*}{\textwidth}{@{\extracolsep{\fill}}ll*{4}{rc}@{}}
\toprule
\multirow{2}{*}{\textbf{Model}} & \multirow{2}{*}{\textbf{System}}
& \multicolumn{2}{c}{\textbf{D1: Linear Cohesion}}
& \multicolumn{2}{c}{\textbf{D2: Argument Coherence}}
& \multicolumn{2}{c}{\textbf{D3: Global Consistency}}
& \multicolumn{2}{c}{\textbf{D4: Overall Quality}} \\
\cmidrule(lr){3-4}\cmidrule(lr){5-6}\cmidrule(lr){7-8}\cmidrule(lr){9-10}
& & \textbf{Avg.} & \textbf{ICC} & \textbf{Avg.} & \textbf{ICC} & \textbf{Avg.} & \textbf{ICC} & \textbf{Avg.} & \textbf{ICC} \\
\midrule
\multirow{2}{*}{deepseek-v3.2} & Baseline & 4.24 (0.76) & 0.997 & 3.49 (0.85) & 0.997 & 3.93 (0.97) & 0.998 & 4.17 (0.70) & 0.992 \\
& \jiajun{LogicTree} & 6.24 (0.47) & 0.994 & 6.26 (0.31) & 0.987 & 6.27 (0.74) & 0.998 & 6.03 (0.59) & 0.991 \\
\addlinespace[2pt]
\multirow{2}{*}{gpt-5.4} & Baseline & 3.26 (0.76) & 0.997 & 2.57 (0.72) & 0.997 & 3.15 (0.96) & 0.998 & 3.32 (0.74) & 0.993 \\
& \jiajun{LogicTree} & 6.65 (0.26) & 0.986 & 6.75 (0.14) & 0.961 & 6.56 (0.41) & 0.994 & 6.66 (0.30) & 0.978 \\
\addlinespace[2pt]
\multirow{2}{*}{glm-5} & Baseline & 4.31 (0.93) & 0.995 & 3.53 (0.88) & 0.994 & 3.62 (0.87) & 0.994 & 4.24 (0.84) & 0.986 \\
& \jiajun{LogicTree} & 6.05 (0.45) & 0.984 & 5.99 (0.42) & 0.980 & 6.08 (0.60) & 0.991 & 5.82 (0.80) & 0.988 \\
\bottomrule
\end{tabular*}
\normalsize
\end{table*}

\subsubsection{General impressions}
Participants consistently \jiajun{described differences between} their writing workflows \jiajun{with LogicTree and} Baseline.
Two primary themes emerged from their feedback, each revealing both strengths and areas for refinement.

\textbf{Structured and controllable top-down writing. }
\jiajun{LogicTree} enabled a structured top-down writing approach aligned with \jiajun{participants' reported} practices.
Participants (10/12) valued starting with a high-level outline to establish a logical skeleton before adding details, providing clear expansion direction. 
As P1 noted, \textit{``It naturally facilitates a step-by-step process, making the overall logic much smoother.''} 
The \jiajun{structure-grounded writing} agent further supported this workflow by allowing users to specify editing scope and review proposed changes before acceptance.
In contrast, Baseline was perceived as encouraging a more linear, bottom-up approach, where users received a dense block of text and had to work backward to infer its underlying logic. 
P9 stated, \textit{``This approach could lead to prematurely getting stuck in the details before the overall structure was sound.''}
However, some participants (3/12) noted that the hierarchical structure, while clarifying, could feel rigid. 
P7 and P9 observed that it forced an early commitment to organization, limiting their ability to explore alternative patterns during drafting. 
Nevertheless, participants appreciated that \jiajun{LogicTree} gave them a clear structural foundation to build upon, with P2 noting, \textit{``Once I commit to the structure, the AI helps me fill it in efficiently, which ultimately saves time.''}

\textbf{Practical logic review and revision. }
\jiajun{LogicTree's structural logical review module helped participants identify and address potential logical issues.}
Beyond structuring the initial draft, the system helped participants inspect local coherence and global consistency through visual error markings. 
P4 appreciated how the visualization of arguments and their associated evidence within the tree structure made it easy to evaluate whether the support was sufficient. 
The \jiajun{structure-grounded writing agent} further assisted by providing explainable suggestions for addressing issues, which participants could accept, reject, or modify.
This contrasted sharply with Baseline, where participants had to read the entire text to find flaws. 
However, reliance on automated error detection raised concerns about uncritical acceptance of system judgments. 
P3 noted, \textit{``While the system suggestions were useful for producing a polished final product, I prefer to think of ways to correct the error myself to improve my thinking skills.''} 
Despite this tension, participants acknowledged that the system lowered the barrier to finding problems and provided a starting point for deeper reasoning. 
As P5 summarized, \textit{``The AI flags issues I might have missed, and then I decide how to address them. It feels like a partner rather than a black box.''}

\section{Discussion}

We reflect on the theoretical foundations of the \textit{object-oriented writing} paradigm, its implications for human-AI interaction with \jiajun{inspectable} logic, and the limitations that point to future work.

\subsection{\jiajun{\textit{Object-Oriented Writing} as a Design Paradigm}}
A central tension in the \textit{object-oriented writing} paradigm lies between providing structural scaffolding for logical reasoning and preserving flexibility. 
\jiajun{The paradigm draws inspiration from object-oriented design, which represents systems as identifiable objects with state, available operations, and explicit relations \cite{booch2008object}.}
\jiajun{OOW borrows this object-centered representational logic without requiring programming-language mechanisms such as classes or inheritance.}
OOW adopts an external representation approach \cite{zhao2026surveylargelanguagemodels, kirsh2010thinking}, making argument structure manipulable rather than merely readable \cite{scaife1996external}. 
\jiajun{In LogicTree, the outline hierarchy serves as what distributed cognition theorists term a cognitive artifact \cite{hutchins1995cognition, hollan2000distributed, rogers2012hci}, reshaping how writers reason about logical structure.}
While participants valued \jiajun{LogicTree's} structured workflow, they noted the hierarchical form could feel rigid and force early commitment to organizational decisions \cite{shipman1999formality}. 
\jiajun{This limitation concerns LogicTree's hierarchical instantiation rather than a requirement that every OOW system use a tree.}
More broadly, it reflects the fundamental representational tradeoff between providing structure that scaffolds reasoning and preserving flexibility for exploratory thinking \cite{kirsh2009problem}. 
\jiajun{The distinction between container and propositional objects partly relieves this tension by letting authors work at multiple granularities, but it does not remove the need to move between structured and linear representations.}

Beyond academic writing, this object-oriented framing suggests a \jiajun{possible design direction} for human-AI interaction \cite{10.1145/302979.303030, shneiderman2020human}. 
For HCI, the approach contributes to feedforward mechanisms \cite{shneiderman1983direct} by making system reasoning inspectable and modifiable before output generation, directly addressing the gulfs of execution and evaluation \cite{norman1986cognitive}. 
In data analysis, users could manipulate analytical steps as objects \cite{heer2012interactive, kandel2012enterprise}. 
In creative design, decisions could be encapsulated as objects with traceable dependencies \cite{grossman2010chronicle}.
In collaborative work, shared artifacts could be decomposed into modular objects with clear ownership \cite{hilliges2007designing}.  
Across these domains, the common thread is shifting from interacting with opaque outputs to manipulating the underlying logic that generates them. 
This vision aligns with emerging HCI research on transparent AI \cite{adadi2018peeking, gunning2019xai} and suggests a future where human-AI collaboration centers on shared manipulable representations of reasoning, making coordination more controllable and aligned with human intent \cite{hoffman1812metrics}.
\jiajun{At the same time, the paradigm assumes that useful aspects of writing can be represented as identifiable objects, an abstraction that may feel unnatural for exploratory or narrative forms like creative nonfiction.}
\jiajun{LogicTree's logical review framework, built on the Toulmin model and cohesion theory, is tailored to argumentative writing and may not transfer directly to genres prioritizing temporal flow or aesthetic expression.}
\jiajun{Our evidence therefore speaks to academic writing rather than to structured knowledge work in general.}
Future work should therefore explore the balance between object-oriented and linear representations and develop genre-adaptive review modules.

\subsection{HAI with \jiajun{Inspectable} Logic}

\jiajun{Making logical criteria machine-readable} introduces a new form of human-AI interaction where formal criteria become inspectable and negotiable rather than binding. 
The structural logical review framework in \jiajun{LogicTree} treats logical quality as \jiajun{a set of LLM-checkable criteria rather than a formally decidable property}, drawing \jiajun{inspiration} from software engineering formal methods \cite{clarke1997model, baier2008principles} and a series of linguistic theories \cite{hasan1984coherence, halliday2014cohesion, toulmin2003uses, rst}. 
By decomposing logical quality into linear cohesion, argument coherence, and global consistency, the system transforms abstract norms into machine readable criteria, enabling automated detection that helps users locate issues without manually reading entire drafts\jiajun{, while the resulting flags remain suggestions rather than guarantees of logical correctness}. 
However, this computational approach has limitations. 
P2 and P12 encountered false positive warnings, reflecting the gap between formal logic and natural argumentation, which echoes critiques of early automated reasoning systems \cite{winograd1972understanding}. 
More fundamentally, there is an inherent tension between making logic computable and respecting the interpretive flexibility of academic writing, where disciplinary conventions and authorial voice resist full formalization \cite{marius1991genre, bazerman1988shaping}. 
P6 also noted that the system's strict detection standard flagged issues that experienced writers considered acceptable, adding overhead to the editing process.

For human-AI interaction, this points toward designing systems where \jiajun{explicit review criteria serve} as a shared referent rather than a binding constraint, enabling contestable AI \cite{alfrink2023contestable}. 
Concrete directions include adaptive logic models that learn from user corrections to reduce false positives, and adjustable detection thresholds users can calibrate to disciplinary context. 
Interfaces can also support conversational contestation where users explain why a flagged passage is acceptable, refining the system's understanding of domain norms. 
These practices shift AI from an authority to a negotiable partner while respecting the interpretive flexibility of academic writing.

\subsection{Limitations and Future Work}
Despite the promising results, our approach has several limitations that point to future research directions.

\textbf{System scalability and interaction fluidity.} 
The object tree view, while effective for structuring logical arguments, faces scalability challenges as manuscripts grow in depth and breadth, potentially overwhelming screen real estate and reducing navigability. 
Additionally, \jiajun{LogicTree} currently processes \jiajun{agent} requests sequentially, forcing users to wait for one \jiajun{request} to finish before editing another independent subtree. 
Future work could address scalability through collapsible cluster nodes that abstract away subtrees, and improve fluidity with parallelized AI processing for independent branches as well as asynchronous background review.


\textbf{Experimental limitations.} 
Our within-subjects study with 12 participants provides initial evidence but limits generalizability. 
The 20-minute tasks may not fully capture authentic writing behaviors that typically unfold over days or weeks, and automated evaluation by LLM judges, while consistent, \jiajun{reflects score stability rather than independently established rating validity}. 
\jiajun{Moreover, comparing the complete LogicTree workflow with a conversational baseline does not isolate the independent effects of the object representation, the writing assistance module, and the structural logical review module; the feature ratings describe perceived usefulness and do not replace an ablation.}
Future work should conduct longitudinal field studies with diverse participants and \jiajun{compare structured baselines and component ablations}.

\section{Conclusion}
\jiajun{We introduced \textit{object-oriented writing}, an OOD-inspired representation and interaction paradigm in which persistent writing objects serve as shared, manipulable referents for authors and AI.}
\jiajun{We instantiated this paradigm in LogicTree, an Overleaf-integrated prototype that distinguishes container and propositional objects and operates structure-grounded writing assistance and structural logical review modules over the same object representation.}
\jiajun{Together, these modules support an iterative authoring loop in which users can inspect the scope of proposed edits, review potential issues in local coherence and global consistency, and retain final control over changes.}
\jiajun{A within-subjects study (N=12) provides initial evidence that the integrated LogicTree workflow reduced workload and improved measured logical quality relative to a conversational baseline in controlled academic writing tasks.}
\jiajun{By making evolving writing structures directly referable and manipulable, this work advances object-centered intermediate representations as a design direction for structured human-AI academic authoring.}

\section*{Generative AI Use Disclosure}
We used OpenAI Codex to assist with writing code for the LogicTree system.
After data collection, we also used Codex to generate supporting Python scripts for aggregating the original data and producing statistical figures, and to polish selected passages of the manuscript.
The authors specified the analytical procedures and reviewed and verified the generated code, aggregated results, statistical figures, and revised text.
The core research ideas, detailed design decisions, key system prompts, non-statistical figures, study design and execution, data collection, and interpretation of the results were led and controlled by the authors.


\bibliographystyle{ACM-Reference-Format}
\bibliography{references}

\end{document}